\documentclass[AMA,STIX1COL]{WileyNJD-v2}
\usepackage{dsfont}
\usepackage[utf8]{inputenc}
\usepackage{amsthm,amsmath}
\usepackage{bm}

\articletype{Discussion}%

\received{26 April 2016}
\revised{6 June 2016}
\accepted{6 June 2016}

\begin{document}

\title{Methodological considerations for novel approaches to covariate-adjusted indirect treatment comparisons}

\author[1]{Antonio Remiro-Az\'ocar}

\author[2,3,4]{Anna Heath}

\author[2]{Gianluca Baio}

\authormark{REMIRO-AZ\'OCAR \textsc{et al}}

\address[1]{\orgdiv{Medical Affairs Statistics}, \orgname{Bayer plc}, \orgaddress{\state{Reading}, \country{United Kingdom}}}

\address[2]{\orgdiv{Department of Statistical Science}, \orgname{University College London}, \orgaddress{\state{London}, \country{United Kingdom}}}

\address[3]{\orgdiv{Child Health Evaluative Sciences}, \orgname{The Hospital for Sick Children}, \orgaddress{\state{Toronto}, \country{Canada}}}

\address[4]{\orgdiv{Dalla Lana School of Public Health}, \orgname{University of Toronto}, \orgaddress{\state{Toronto}, \country{Canada}}}

\corres{*Antonio Remiro-Az\'ocar, Medical Affairs Statistics, Bayer plc, Reading, United Kingdom. \email{antonio.remiro-azocar@bayer.com}. Tel: (+44) 7818 746 781} 

\presentaddress{Antonio Remiro-Az\'ocar, Medical Affairs Statistics, Bayer plc, 400 South Oak Way, Reading, RG2 6AD, United Kingdom}

\abstract{
We examine four important considerations in the development of covariate adjustment methodologies for indirect treatment comparisons. Firstly, we consider potential advantages of weighting versus outcome modeling, placing focus on bias-robustness. Secondly, we outline why model-based extrapolation may be required and useful, in the specific context of indirect treatment comparisons with limited overlap. Thirdly, we describe challenges for covariate adjustment based on data-adaptive outcome modeling. Finally, we offer further perspectives on the promise of doubly-robust covariate adjustment frameworks. 
}

\keywords{Health technology assessment, indirect treatment comparison, matching-adjusted indirect comparison, parametric G-computation, covariate adjustment, model misspecification}

\maketitle

\renewcommand{\thefootnote}{\alph{footnote}}

Covariate-adjusted indirect comparisons are increasingly used in health technology assessment (HTA) to compare treatments that have not been trialed against each other.\cite{phillippo2016nice, phillippo2018methods} Matching-adjusted indirect comparison (MAIC)\cite{signorovitch2010comparative} is the most popular methodology.\cite{phillippo2019population} MAIC weights the subjects in one of the studies by the odds of being assigned to another study, conditional on a set of baseline covariates, without explicitly requiring an outcome model. The estimated weights enforce balance in selected covariate moments between studies. 

Recently, alternatives based on estimating the conditional outcome expectation given treatment and baseline covariates have been proposed. In an article published in this journal, we develop an adaptation of parametric G-computation.\cite{remiro2022parametric} This involves fitting a parametric outcome model to patient-level data for one of the studies to predict and contrast potential outcomes in another study. When assumptions hold, parametric outcome modeling offers greater statistical precision and efficiency than MAIC.\cite{remiro2022parametric} Nevertheless, this does not imply that the former is inherently superior to the latter.  

Where there is limited overlap between covariate distributions across studies, outcome modeling methods extrapolate the association between covariates and outcome. In an insightful commentary, Vo warns of the dangers of extrapolating a misspecified outcome model.\cite{vo2023cautionary} Valid model-based extrapolation depends on strong assumptions and relies on accurately capturing the true covariate-outcome relationships. Vo advocates for the use of MAIC over G-computation because the former does not extrapolate, and outlines the pitfalls of using data-adaptive methods to estimate the outcome model.\cite{vo2023cautionary}  

We congratulate Vo on his excellent editorial and thank the Editor for arranging this exchange. In this counter-response, we examine four important considerations for the development of covariate adjustment methodologies in the context of indirect treatment comparisons. Firstly, we consider potential advantages of weighting versus outcome modeling, placing focus on bias-robustness. Secondly, we outline why model-based extrapolation may be required and useful, in the specific context of indirect treatment comparisons with limited overlap. Thirdly, we describe challenges for covariate adjustment based on data-adaptive outcome modeling. Finally, we offer further perspectives on the promise of doubly-robust covariate adjustment frameworks. 

\section{Bias-robustness: the case for weighting}\label{sec1}

Insofar, covariate-adjusted indirect comparison methods have relied on parametric assumptions for unbiased estimation. In practice, these assumptions will fail to some extent. Therefore, we must think about the bias-robustness of different approaches. In parametric G-computation, bias arising from model misspecification cannot be easily verified in the data. As highlighted by Vo in his first simulated example, an outcome model that seems approximately correct when fitted to the patient-level data would not necessarily fit well in extrapolated regions.\cite{vo2023cautionary} In practice, candidate models are compared using diagnostic plots, information criteria or predictive performance measures such as the mean squared prediction error. These measures cannot detect model misspecification in extrapolated regions. Most model performance measures average over the observed covariate distribution, and are not necessarily transportable between different covariate distributions.\cite{steingrimsson2023transporting} 

In weighting approaches, evaluating the reduction in bias due to covariate adjustment is more straightforward. Weighting methods do not necessarily require knowledge of the trial assignment process. They seek weights that attain cross-study balance in the distributions of measured covariates. By checking for adequate balance after weighting, one can assess whether bias due to differences in observed covariates has been mitigated. 

Vo suggests adopting the individual-level covariate generation step proposed by Remiro-Az\'ocar et al.\cite{remiro2022parametric} to fit the trial assignment model, typically a logistic regression, using maximum-likelihood estimation.\cite{vo2023cautionary} This approach explicitly models the conditional probability of trial assignment as a function of the baseline covariates, with odds weights derived from the estimated probabilities.\cite{westreich2017transportability} The typical model-building strategy is outcome-blind and evaluates different specifications on a trial-and-error basis. One may begin with a model containing additive terms for the relevant covariates, and consider including non-linear transformations or higher-order terms, e.g.~polynomials, interactions, if imbalances remain.\cite{kyle2019evaluating} 

MAIC follows a different paradigm for weighting. It directly enforces balance in covariate moments without explicitly modeling the conditional probability of trial assignment. Covariate balance is viewed as a convex optimization problem and the weights are solutions to such problem. MAIC is an entropy balancing technique.\cite{josey2021transporting, phillippo2020equivalence} Entropy balancing approaches are more stable, efficient and robust to model specification than the standard ``inverse weighting'' modeling approaches.\cite{josey2021transporting, cheng2022double} 

Recent results imply that MAIC, balancing for covariate means, is asymptotically unbiased if covariates have a linear relationship with the mean difference (in the anchored scenario) or the absolute outcome means (in the unanchored scenario), even if the implicit trial assignment model is incorrectly specified.\cite{josey2021transporting, cheng2022double} Nevertheless, MAIC may be subject to bias if there are non-linearities that are not accounted for by the, also implicit, outcome model. MAIC remains biased in Vo’s first simulated example, partly because there is a squared covariate-by-treatment product term in the outcome-generating model, and only first-order moments (sample means) and not second-order moments (sample variances) are balanced for.\cite{vo2023cautionary} Because second-order balance is enforced by balancing the means of the squared covariates,\cite{phillippo2016nice} balancing both first- and second-order moments would have provided some protection against bias. However, there is a risk that MAIC is unable to find a solution to the convex optimization problem, and there may also be a substantial cost in terms of precision.\cite{campbell2023standardization}

In summary, model misspecification bias is not easily diagnosed for outcome modelling, particularly when there is extrapolation. Conversely, balance diagnostics for weighting offer some indication on the level of bias reduction due to covariate adjustment. It is important to make a distinction between modeling and balancing approaches to weighting. Generally, the latter are less susceptible to bias and offer greater stability and precision than the former.

\section{Model-based extrapolation: undesirable but necessary}\label{sec2}

Where overlap is poor, weighting is inefficient. Effective sample sizes are low and results may be sensitive to the inordinate influence of a few extreme weights. A drop in precision is expected for all covariate adjustment methods as overlap decreases. In our original simulation study, such reduction is more marked for MAIC than for parametric G-computation.\cite{remiro2022parametric} When modeling assumptions hold, the latter is more precise and efficient than the former across all scenarios.\cite{remiro2022parametric} The increase in precision is achieved by extrapolating into non-overlapping regions of the covariate space. The extrapolation is made using strong parametric assumptions, which may be unfeasible. In our original simulation study, variance estimates and coverage rates are valid for G-computation with a correctly specified outcome model.\cite{remiro2022parametric} Even so, the interval estimates do not typically reflect the extrapolation uncertainty. This could be problematic when modeling assumptions fail, particularly given that the appropriate characterization of uncertainty is a central objective of HTA.\cite{claxton2005probabilistic}

Vo argues that weighting approaches reflect uncertainty more honestly.\cite{vo2023cautionary} This point is also raised in the discussion of our article,\cite{remiro2022parametric} following Vansteelandt and Keiding.\cite{vansteelandt2011invited} The degree of extrapolation made by parametric G-computation is implicit; it cannot be controlled and may hide any underlying lack of overlap. Conversely, the presence of extreme weights in MAIC explicitly manifests the high estimation uncertainty in data-sparse regions. 

Vo recommends reducing the impact of extreme weights without resorting to extrapolation.\cite{vo2023cautionary} A suggested approach is weight truncation. Remiro-Az\'{o}car has explored truncation in combination with MAIC, and shown that it can produce substantial precision and efficiency improvements when sample sizes are low and the extremity of the weights is high.\cite{remiro2022two} However, truncation induces asymptotic bias and is prone to ad hoc heuristics, e.g.~deciding on cutoff thresholds.\cite{remiro2022two} Given the arbitrary nature of such decisions, weight truncation is arguably not ``honest'' in quantifying uncertainty. 

Weight stabilization is also recommended by Vo.\cite{vo2023cautionary} The version of MAIC in our original article fits a weighted marginal model of outcome on a time-fixed binary treatment.\cite{remiro2022parametric} Because this model is ``saturated'' --- there is a parameter for each unique predictor combination ---- stabilized and unstabilized weights yield identical results.\cite{shiba2021using} Jackson et al. have developed an adaptation to MAIC that minimizes the dispersion of the weights, improving precision but inducing some bias.\cite{jackson2021alternative} Remiro-Az\'{o}car recently proposes a modular extension to MAIC, which improves precision and efficiency without resorting to bias-variance trade-offs, particularly when sample sizes are low and covariates are highly prognostic.\cite{remiro2022two} It does so by estimating the treatment assignment mechanism in the study with patient-level data, and is therefore not applicable in the unanchored case. 

The aforementioned approaches are useful for variance reduction, yet depend on weighting solutions being available in the first place. Where there is no common support, feasible weighting solutions providing satisfactory balance may not exist.\cite{phillippo2020assessing} If the observed range of a covariate in the study with patient-level data does not cover the respective mean published for the comparator study, MAIC suffers from convergence failures and cannot even produce an estimate.\cite{jackson2021alternative} According to Vo, ``we might need to accept that the difference between populations is too large to be adjustable''.\cite{vo2023cautionary} Nevertheless, treatment effect estimates are needed for HTA and, in many jurisdictions, to populate the economic models required to make reimbursement decisions. Vo views model-based extrapolation as a flaw.\cite{vo2023cautionary} In our view, extrapolation beyond the overlapping regions of the covariate space may be necessary when the study samples are divergent. 

Finally, in the anchored scenario, Vo suggests excluding covariates that overlap poorly but are not effect modifiers from the MAIC adjustment,\cite{vo2023cautionary} as indicated by current guidance.\cite{phillippo2018methods} Our simulation study assumes a best-case situation where these covariates can be identified. However, this is challenging in practice, due to the scale dependence of effect modification and to limited knowledge on the drivers of treatment effect heterogeneity, particularly for novel therapies. As highlighted by Vo in his second simulated example, excluding weak effect modifiers in order to reduce standard errors may induce bias.\cite{vo2023cautionary} There is a tension between satisfying the conditional constancy or transportability of effects across studies and maintaining overlap. Accounting for a greater number of covariates increases the plausibility of the former but decreases the likelihood of the latter. 

In conclusion, parametric G-computation relies on model-based extrapolation to overcome limited overlap and improve precision with respect to MAIC. Model-based extrapolation depends on strong assumptions and is typically viewed as an undesirable feature. Nevertheless, it may be necessary in practice to generate the treatment effect estimates that are required for HTA decision-making.

\section{The perils of data-adaptive G-computation}\label{sec3}

For asymptotically unbiased estimation, parametric G-computation relies on the correct specification of the outcome model. Parametric models impose restrictive assumptions on functional forms; for instance, that effects are linear and additive on some transformation of the conditional outcome mean. These assumptions may not be plausible where there are multiple continuous covariates and complex non-linear relationships, particularly if background theory is weak. Making strong parametric assumptions has a cost. If the parametric model is incorrectly specified, the G-computation estimator is subject to bias and this bias does not decrease with sample size, at any rate. This may lead to substantial undercoverage: as the sample size grows, the probability that the interval estimates contain the target estimand shrinks to zero. 

To reduce the risk of bias resulting from parametric model misspecification, one may consider using data-adaptive methods, e.g. non-parametric techniques or machine learning, to estimate the conditional outcome expectations.\cite{le2021g} These approaches make weaker modeling assumptions, but the flexibility also comes at a cost. From a frequentist perspective, that is a slower rate of convergence than the parametric rate.\cite{naimi2017challenges} A misspecified parametric estimator attains fast $n^{1/2}$ convergence, but to the wrong target value. Data-adaptive estimators are more likely to converge to the true expectations, but converge more slowly. This may lead to larger-than-desirable bias and poor coverage in finite samples. 

Consider the naive use of lasso-based G-computation in the setting explored by Vo.\cite{vo2023cautionary} The lasso is a regularization estimator that trades off variance for bias to avoid overfitting and improve out-of-sample predictive performance. This results in slow convergence rates, with bias and mean square error diminishing slowly with sample size, particularly in high-dimensional settings. In Vo's second simulated example,\cite{vo2023cautionary} the lasso dismisses various covariates that are weak effect modifiers but are correlated with study assignment, ultimately leading to bias.  

In addition to slower than parametric convergence rates, data-adaptive G-computation estimators also have non-normal sampling distributions.\cite{van2014higher} When using parametric outcome models, the central limit theorem and the law of large numbers can be applied to obtain convenient asymptotic approximations. This enables one to readily construct variance and interval estimates, e.g.~through resampling methods such as the bootstrap and Wald normal interval equations. Conversely, the asymptotic distribution of data-adaptive estimators often is irregular and cannot be derived. This makes uncertainty quantification difficult. Even where one can establish the asymptotic distribution, the use of the bootstrap to compute interval estimates is not supported by theoretical results as it requires certain convergence rate conditions to hold.\cite{bickel2012resampling}

We propose moving away from a frequentist framework and exploring the use of Bayesian G-computation. Non-parametric outcome models such as Bayesian additive regression trees (BART)\cite{chipman2007bayesian, hill2011bayesian, hahn2020bayesian} are very flexible and well-suited to capture complex functional forms. Because BART directly samples from the posterior distribution of the treatment effect, principled variance and interval estimation are possible without having to resort to the bootstrap. Bayesian methods naturally integrate the analysis into a probabilistic formulation that is desirable for HTA,\cite{claxton2005probabilistic} and allow the incorporation of substantive prior information. 

BART is implemented in many software packages, requires little parameter tuning, and has performed excellently in causal inference applications.\cite{hill2011bayesian, kern2016assessing} Nevertheless, BART is not a panacea. As per the lasso, it regularizes heavily to penalize overfitting, which may induce non-negligible bias into the estimation of treatment effects.\cite{hahn2020bayesian} Little is known theoretically about BART's asymptotic (frequentist) properties. In the causal inference literature, the method has exhibited poor coverage in scenarios with limited overlap because it does not extrapolate well.\cite{zhu2023addressing} Regression trees are piecewise step functions that partition the covariate space into disjoint cuts. Consequently, they extrapolate beyond the overlap region using a constant value.\cite{zhu2023addressing}

In conclusion, the blind use of data-adaptive estimation together with G-computation may lead to bias in finite samples and is constrained by limited theoretical justification for valid statistical inference. Data-adaptive G-computation estimators are more flexible than their parametric counterparts. Nevertheless, flexible estimators still extrapolate where overlap is limited. The extrapolations of very flexible models may be intrinsically flawed, impairing point and interval estimation under poor overlap. 

\section{The appeal of doubly-robust adjustment}\label{sec4}

Vo proposes the use of doubly-robust approaches such as augmented weighting estimators,\cite{vo2023cautionary} in conjunction with the individual-level covariate simulation step by Remiro-Az\'ocar et al.\cite{remiro2022parametric} Augmented weighting methods apply two working models: a model for trial assignment conditional on covariates, and another for the outcome conditional on treatment and covariates. Only one of the two models needs to be correctly specified to achieve asymptotically unbiased estimation. In general, doubly-robust estimators should be less prone to model misspecification bias than singly-robust estimators; they offer two opportunities for valid adjustment instead of one, mitigating the risks of applying the wrong outcome model. Nevertheless, the two working models are typically parametric. In practice, there is seldom adequate background knowledge on the true trial assignment and outcome-generating processes. Therefore, both parametric models may be incorrect, in which case doubly-robust adjustment is subject to bias. 

A valuable suggestion by Vo is the application of data-adaptive methods within a doubly-robust framework.\cite{vo2023cautionary} An oft-ignored advantage of such framework is that it allows for the slower convergence of working models.\cite{chernozhukov2018double, zeng2023efficient} Augmented weighting approaches combine two singly-robust estimators, such that the overall convergence rate is as fast or faster than the convergence rate of each separate estimator. Consider that a flexible non-parametric regression or machine learning technique is used to fit both working models. Even if each working model converges at a rate slower than the optimal parametric $n^{1/2}$ rate, the product of the individual rates may still be $n^{1/2}$ or faster.\cite{chernozhukov2018double, zeng2023efficient} Cross-validation based ensembles of data-adaptive techniques (e.g.~Super Learner) also satisfy the rate condition if one of the candidate estimators does.\cite{vaart2006oracle}

Valid inference for data-adaptive doubly-robust estimators remains a challenge. Their asymptotic normality, which allows for simple approaches to construct variance and interval estimates, relies on an unverifiable ``Donsker'' condition from empirical process theory.\cite{van1996weak} Such condition may be inappropriate in high-dimensional settings, restricting the complexity of the data-adaptive methods that are permitted. 

The Donsker condition can be relaxed using sample-splitting procedures such as cross-fitting.\cite{chernozhukov2018double, zivich2021machine} These create an additional source of variation in finite samples, and need to be iterated repeatedly to remove seed-dependence.\cite{chernozhukov2018double, zivich2021machine} This is computationally burdensome if there is no closed-form expression for the variance, as bootstrapping would incur considerable run-time. The asymptotic distribution of estimators can be learned by characterizing their efficient influence function, as suggested by Vo.\cite{vo2023cautionary} Nevertheless, we suspect that finding tractable approximations is not trivial with limited access to patient-level data. 

In summary, doubly-robust approaches require the correct specification of either a trial assignment or an outcome model to obtain consistent estimation. While singly-robust data-adaptive estimators may suffer from slow convergence rates, doubly-robust estimators can accommodate the use of slower-converging data-adaptive models. Nevertheless, their asymptotic normality relies on an unverifiable Donsker condition, which restricts the complexity of the models that are permitted. Such condition could be overcome through the use of sample-splitting, but this may require substantial computational resources when closed-form expressions for the variance are unavailable. 


\section{Conclusions}\label{sec5}

Weighting is the most popular approach for covariate adjustment in the context of indirect treatment comparisons with limited patient-level data. Recently, alternatives based on outcome modeling have been developed. Such methods enable the estimation of treatment effects in ``limited overlap'' settings where MAIC cannot be applied. Under no failures of assumptions, parametric outcome modeling increases precision and efficiency with respect to weighting, particularly when overlap is poor. It does so by extrapolating outside the overlap region, which involves untestable and potentially untenable parametric assumptions. While functional forms may hold approximately in the overlap region, this is not necessarily the case when extrapolating beyond it.  

Specifying an appropriate outcome model -- one that seems approximately correct in the overlap region but that also extrapolates well -- is challenging in many settings, and can quickly become an overwhelming process. Analysts are typically drawn to simple but restrictive models, assuming linear effects to be the norm. Subject-matter knowledge about the outcome-generating process is often fragile. Data-adaptive outcome modeling techniques are attractive to mitigate the risk of bias arising from parametric model misspecification. These may not require the pre-specification of functional forms and can provide some automation to the model selection process. Nevertheless, the slow convergence rates of flexible models may also lead to bias. In addition, flexible estimators can extrapolate poorly and may offer limited theoretical justification for valid uncertainty propagation. 

Precision and efficiency under no failures of assumptions should not be the only characteristics determining the choice of a suitable estimator. In general, the level of bias reduction due to covariate adjustment is easier to assess for weighting than for outcome modeling. Several authors argue that entropy balancing approaches to weighting such as MAIC are more bias-robust than outcome modeling methods.\cite{vo2023cautionary, josey2021transporting, cheng2022double} Doubly-robust methods that combine a trial assignment model with a model for the conditional outcome expectation offer two opportunities for correct model specification, and are likely more bias-robust than both MAIC and parametric G-computation. The doubly-robust framework also allows for the use of slower-converging data-adaptive models, although adequate variance and interval estimation may have its challenges. 

Standard (unadjusted) indirect treatment comparisons rely on a very strong assumption: the unconditional constancy or transportability of relative effects across studies. Covariate-adjusted indirect comparisons seek to relax this assumption by adjusting for baseline covariates. In doing so, they rely on the correct specification of parametric models. The selection of an appropriate model is challenging and subject to many ``researcher degrees of freedom''. The development of innovative covariate adjustment methods that help to reduce these degrees of freedom is imperative. The valid characterization of uncertainty, a key requirement of HTA,\cite{claxton2005probabilistic} is another important factor in determining the suitability of these methods for decision-making. The performance of novel covariate adjustment methods should be examined in simulation studies reflecting typical scenarios in the context of indirect treatment comparisons with limited patient-level data. 

Finally, innovative methodologies may not overcome aspects such as limited overlap and the unavailability (or inconsistent definition) of influential covariates. The former may require model-based extrapolation, which always calls for a leap of faith. The latter can be addressed through the development of core patient characteristic sets that define important prognostic covariates to be measured and reported among specific therapeutic areas.\cite{vo2023development}

\section*{Acknowledgments}

The authors thank Tat-Thang Vo for his insightful editorial and the editors of Research Synthesis Methods for facilitating this discussion and inviting us to reply. Anna Heath is funded by Canada Research Chair in Statistical Trial Design; Natural Sciences and Engineering Research Council of Canada (award No. RGPIN-2021-03366).

\subsection*{Financial disclosure}

Funding agreements ensure the authors’ independence in writing and publishing the article.

\subsection*{Conflict of interest}

Antonio Remiro-Az\'{o}car is employed by Bayer plc but declares no conflicts of interest. The views expressed in this article do not necessarily represent those of Bayer plc. Anna Heath and Gianluca Baio declare no potential conflicts of interest. 

\subsection*{Data Availability Statement}

Data sharing is not applicable to this article as no datasets were generated or analyzed.

\bibliography{wileyNJD-AMA}


\end{document}